\documentclass{iopart}

\usepackage{graphicx}
\usepackage{float}
\usepackage{color}

\begin{document}

  \bibliographystyle{unsrt}
  \title [Transitions between epitaxial growth regimes: A (1+1)-dimensional KMC study]{Transitions between epitaxial growth regimes: A (1+1)-dimensional kinetic Monte Carlo study}
  \author{C. S. Dias}
   \ead{\mailto{cristovao@fisica.uminho.pt}}
    \address{GCEP-Centro de F\'isica da Universidade do Minho, 4710-057 Braga, Portugal}

  \author{N. A. M. Ara\'ujo}
   \ead{\mailto{nuno@ethz.ch}}
    \address{Computational Physics for Engineering Materials, IfB, ETH Z\"{u}rich, Schafmattstr. 6, CH-8093 Z\"{u}rich, Switzerland}

  \author{A. Cadilhe}
   \ead{\mailto{cadilhe@lanl.gov}}
    \address{Theoretical Division, MSK 717, Los Alamos National Laboratory, Los Alamos, NM 87545, USA}

  \begin{abstract}
To study epitaxial thin-film growth, a new model is introduced and extensive kinetic Monte Carlo simulations performed for a wide range of fluxes and temperatures. Varying the deposition conditions, a rich growth diagram is found. The model also reproduces several known regimes and in the limit of low particle mobility a new regime is defined. Finally, a relation is postulated between the temperatures of the kinetic and thermal roughening transitions.
  \end{abstract}

  \maketitle

\section{Introduction}

The development of experimental techniques to probe and visualize at the atomic scale has challenged studies on film morphology and the identification of the relevant controlling parameters. Several experimental \cite{Thiel2004,Rosenfeld1995a,Rosenfeld1995,Gyure1998} and theoretical \cite{Evans2006,Michely2004,Pimpinelli1998,Lapujoulade1994} studies have been performed to identify the underlying relevant mechanisms and processes at the atomistic scale. Herein, we introduce a model for multilayer growth to study the influence of deposition conditions, namely, temperature and deposition flux.

Growth is an intrinsically nonequilibrium problem where each deposition event perturbs the system and keeps it out of equilibrium. It competes with thermally activated relaxation events that lead to equilibrium. Particles can arrive and stick to the substrate at a certain flux or hop between different basins, in the energy landscape, through thermally activated processes. Two different characteristic times are of relevance: the typical time between consecutive deposition events, which is related to the incoming flux of particles, and the hierarchy of relaxation times, related to temperature. The ratio between the first and the latter measures the departure from equilibrium \cite{Xiao1994}.

During homoepitaxial growth, where deposited and substrate particles are the same, at temperatures below the roughness transition, the layer-by-layer growth regime is expected with nucleation of subsequent layers occurring after the previous one had time to complete. However, if the time between consecutive deposition events is short, on average, as compared to the typical timescale for the rate determining processes leading to relaxation, then thermal equilibrium is disrupted. Growth takes place away from equilibrium since thermal processes leading to relaxation remain unaltered. For example, one can expect island formation on top of existing islands, therefore leading to the nucleation of new layers before existing layers had a chance to become complete.

We are primarily interested in the cooperative behavior that is established at long timescales and large length scales. We, therefore, depart from techniques that provide a more refined evolution at much shorter timescales and/or length scales \cite{Vvedensky2001,DasSarma1987,DasSarma1990}. To this end we adopt a kinetic Monte Carlo approach, where the list of possible processes is identified \textit{a priori} \cite{Yip2005b, Clarke1987, Rockett1988, Kobayashi1988, Kawamura1989}. In this context, the properties of the model reflect the, \textit{a priori}, chosen set of processes. Since this set of processes is known, it is easier to gain insight on the relevant mechanisms.

There has been a substantial amount of effort focused on the study of submonolayer deposition, island formation and growth, and island-island coalescence \cite{Evans2006,Stoldt1999,Stoldt1998b,Araujo2010,Araujo2008,Cadilhe2007}, while for multilayer growth several questions remain open \cite{Ganapathy2010,Einstein2010,Forgerini2010,Lam2010,Misbah2010}. Typically, either a detailed approach is considered, which attempts to make no assumptions, or as few as possible \cite{Xiao1994, Smilauer1995, Bartelt1995, Landau1996, Meng1996, Smilauer1993, Liu2005}, or simple models are developed, where mechanisms are restricted to the most relevant ones \cite{DasSarma1991, Phillips1991, Dassarma1996, Tamborenea1993, Schimschak1995, Krug1993, AaraoReis2010, Kanjanaput2010}. In this work, we adopt the latter approach, which allows us to better understand the relevance of each process. To rely on as few as possible parameters, we introduce a model where the activation barrier of each process is computed based on the one for the terrace diffusion and the coordination number in the initial and final states. This accounts for well known phenomena like, e.g., the \textit{corner rounding} barrier. We acknowledge that the neighbor counting approach has known limitations, since pair-wise interactions are considered, thus neglecting many body effects. Despite the shortcomings, these models have successfully explained relevant phenomena by addressing length and time scales of the order of the ones considered in this work \cite{Gyure1998,Kaneko1995,Baskaran2011}.

Exploring the two-parameter space of temperature and deposition flux we not only reproduce several growth regimes previously reported, but we also identify a new one, namely, at low particle mobility. In this new regime, the typical analysis based on surface properties is not able to characterize it. Besides, the transition between the various identified regimes is also studied.

The manuscript is organized in the following way. The model is introduced in the next section and simulation results introduced in section~\ref{sec:R_and_D} along with their discussion. Concluding remarks are drawn in section~\ref{conclusions}.

\section{Model and definitions}\label{model}

To study epitaxial growth we propose a model where particles are randomly deposited on the substrate, while deposited particles can then hop in the energy landscape between adjacent basins through thermally activated processes. The set of possible processes on the surface is given by the lattice sites surrounding the initial and final neighboring sites for a total of $2^{10}$ configurations along each of the four possible directions, namely, up, down, left, and right. The activation barrier for each of the possible configurations is simply taken as the difference between the coordination number in the initial and final positions \cite{Smilauer1993}, 
     \begin{equation}
 E_{i}=E_m- (n_f - n_0)J_{pp}, 
     \label{eq:active_energy}
     \end{equation} 
where $E_m$ is the activation barrier for terrace diffusion, $J_{pp}$ the energy of the particle-particle interaction, and the coordination number before (after) the hop is $n_0$ ($n_f$). In spite of its simplicity, this strategy has been able to grasp relevant phenomena like, e.g., the high vacancy mobility observed in several experiments \cite{Poelsema1985}, and the relevance of perimeter mass transport for submonolayer island evolution on Ag(100) substrates \cite{Stoldt1998b}. The rate of process $i$ is obtained from the Arrhenius equation,
     \begin{equation}
        r_i=\nu_i e^{-E_{i}/{k_BT}},
        \label{eq:arrhenius}
      \end{equation} 
where $\nu_i$ is the attempt frequency. For simplicity the same attempt frequency is assumed equal for all processes, so $\nu_i\equiv\nu$ \cite{Voter1984}.

\begin{figure}[t]
     \begin{center}
     \includegraphics[width=6.0cm]{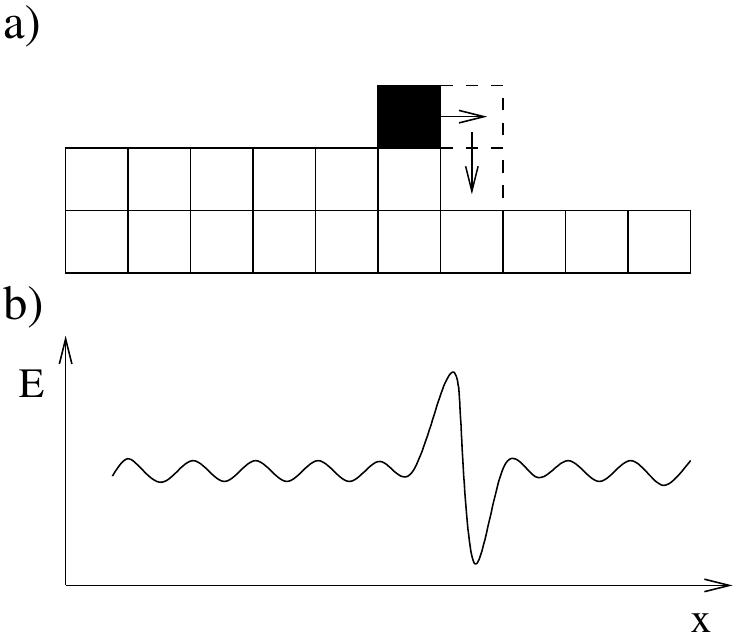}\\
     \end{center}
      \caption{a) Scheme of the \textit{corner rounding} process and of the b) energy landscape.}
      \label{fig:corner_rounding}
    \end{figure}

A key process in this system is the \textit{corner rounding} \cite{Stoldt1998, Cadilhe2000} as illustrated in figure~\ref{fig:corner_rounding}. This process is usually included by considering a direct jump to the step below, which requires an additional parameter for the activation barrier \cite{Bartelt1995, Stoldt1998, Villain1991, Stoldt2000}. In the present model, the rules naturally account for such barrier, with a two steps process.

The ballistic rule for deposition implies that deposited particles stick to the first occupied (nearest) neighbor, so particles can overhang \cite{DasSarma1994,Marmorkos1992,Marmorkos1990}. However, the ballistic deposition model produces too many vacancies. Other models (see for example Ref.~\cite{Schimschak1995, Lanczycki1994}), such as solid-on-solid (SOS) and downward funneling mechanism, have been considered for deposition \cite{Stoldt2000}, but they are not able to reproduce the experimentally observed concentration of vacancies.

For epitaxial growth, particle detachment from the substrate can be neglected \cite{Neave1985}. To achieve such a constraint, processes which break connectivity to the substrate are forbidden. To identify these situations we consider the Hoshen-Kopelman algorithm \cite{Hoshen1976}. This rule prevents clusters from detaching from the substrate, a mechanism that can be neglected for a large region of the diagram of flux and temperature, and it its usefulness is solely restricted to the prevention of unphysical processes common on (KMC) lattice models.

We introduce $\phi$ as the ratio between the binding and thermal energies,
  \begin{equation}
   \phi =\frac{J_{pp}}{k_BT}.
   \label{eq.phi}
  \end{equation} 
The attempt frequency for the terrace diffusion process $\nu_{d}$ is defined as
  \begin{equation}
    \nu_{d}=\nu e^{-\frac{E_m}{k_BT}}.
    \label{eq.D}
  \end{equation} 
Now, if we also define the ratio, $R_E=E_m/J_{pp}$, i.e., between the activation energy for terrace diffusion, $E_m$, and the pair-wise interaction between particles, $J_{pp}$, we can rewrite equation~(\ref{eq.D}) as
  \begin{equation}
    \nu_{d}=\nu e^{-R_E\phi}.
    \label{eq.D2}
  \end{equation} 
According to equations~(\ref{eq:arrhenius})~and~(\ref{eq.D2}), the rate of a certain process $i$ can be calculated by,
   \begin{equation}
      r_i=\nu_{d} e^{\Delta n_i\phi}.
      \label{eq:arrhenius_final}
   \end{equation}
We have casted the set of parameters of the model in terms of two, related to energy ratios, namely, $R_E$ and $\phi$. The former measures the activation barrier of the terrace diffusion in units of the particle-particle interaction while the latter provides a measure of mobility. The last parameter, the flux, $F$, defined as the number of impinging atoms arriving, per unit of surface (a line in the present case) and per unit of time, in monolayers per second (ML/s). 

In the next section we identify and analyze different growth regimes obtained with the present model and discuss their transitions.

\section{Results and discussion}\label{sec:R_and_D}

To reproduce growth regimes, we performed extensive kinetic Monte Carlo simulations of the (1+1)-dimensional system. The time increment is computed as $\Delta t=-\ln{\xi}/\sum_i{r_i}$, where $\xi$ is an uniformly distributed variable in the range (0, 1] and the sum is over all processes accessible to each of the particles, thus giving the total rate of the system. Simulations were carried out with periodic boundary conditions in the horizontal direction. Unless otherwise stated, simulations were performed on a lattice of $10^3$ sites and results were averaged over $10^3$ samples.

We now start with the analysis of the effect of the flux of impinging particles in the morphology of the system. We compute the time evolution of the roughness defined as,

  \begin{equation}
     W(t)=\sqrt{\langle\left[h(x,t)-\langle h(t)\rangle\right]^2\rangle},
     \label{eq:roughness}
  \end{equation} 
where $h(x,t)$ is the height of the column $x$ at time $t$ and $\langle h(t)\rangle$ is the average height at time $t$.

\begin{figure}[t]
\begin{center}
\begin{tabular}{cc}
    \includegraphics[width=6.0cm]{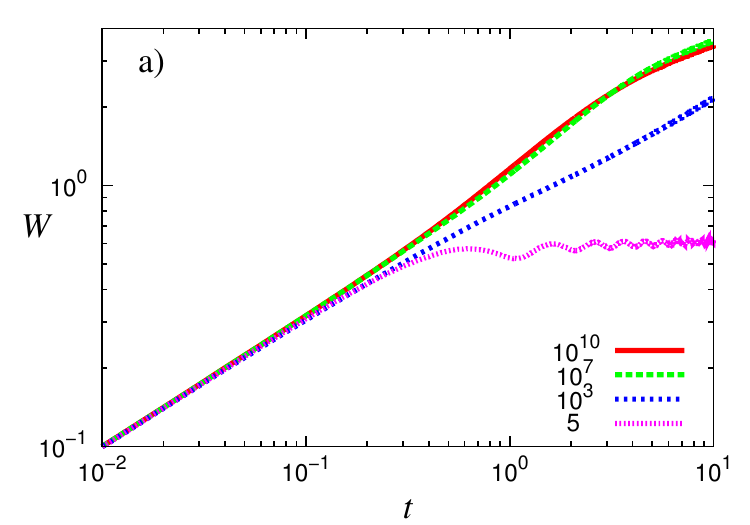} & \includegraphics[width=6.0cm]{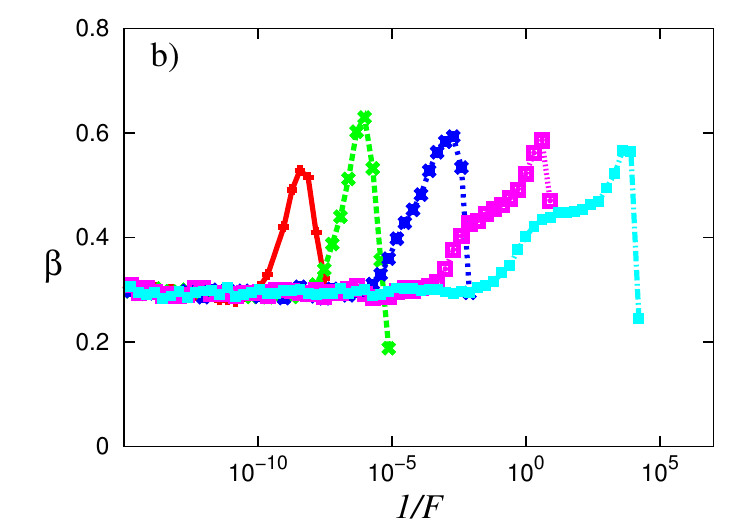}\\
\end{tabular}
    \caption{a) Roughness evolution at $\phi=3$ as a function of time defined in units of deposited layers, for $F=\{5,10^{3},10^{7},10^{10}\}$ML/s. At higher fluxes the growth regime is rough and at lower fluxes the roughness oscillates representing a layer-by-layer growth. b) The roughness exponent $\beta$ as a function of $1/F$ for different values of $\phi$. From left to right $\phi=\{1\mbox{(red)},2\mbox{(green)},3\mbox{(blue)},4\mbox{(violet)},5\mbox{(cyan)}\}$, measurements performed after $10^2$ deposited monolayers.}
    \label{fig:roughness_exponent}
\end{center}
  \end{figure}

\begin{figure}[ht]
 \begin{center}
\begin{tabular}{ccccc}
 \includegraphics[width=3cm]{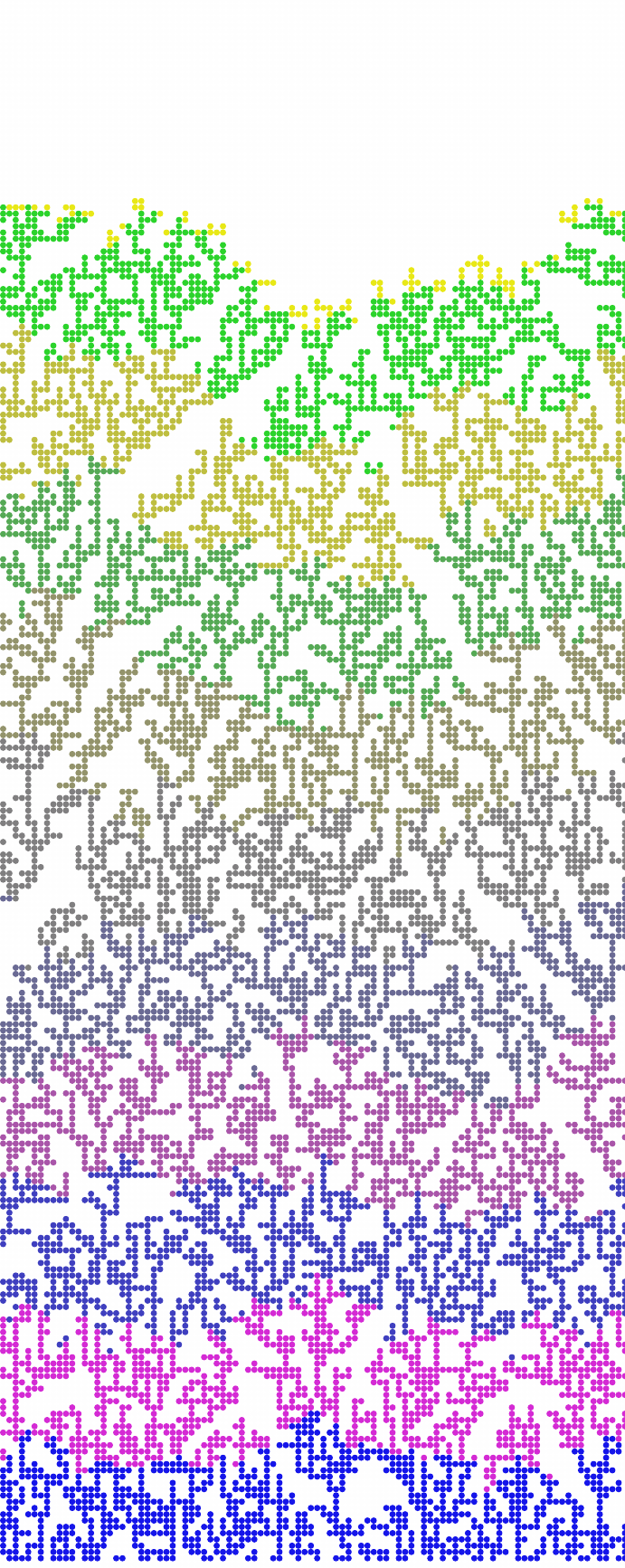} & \includegraphics[width=3cm]{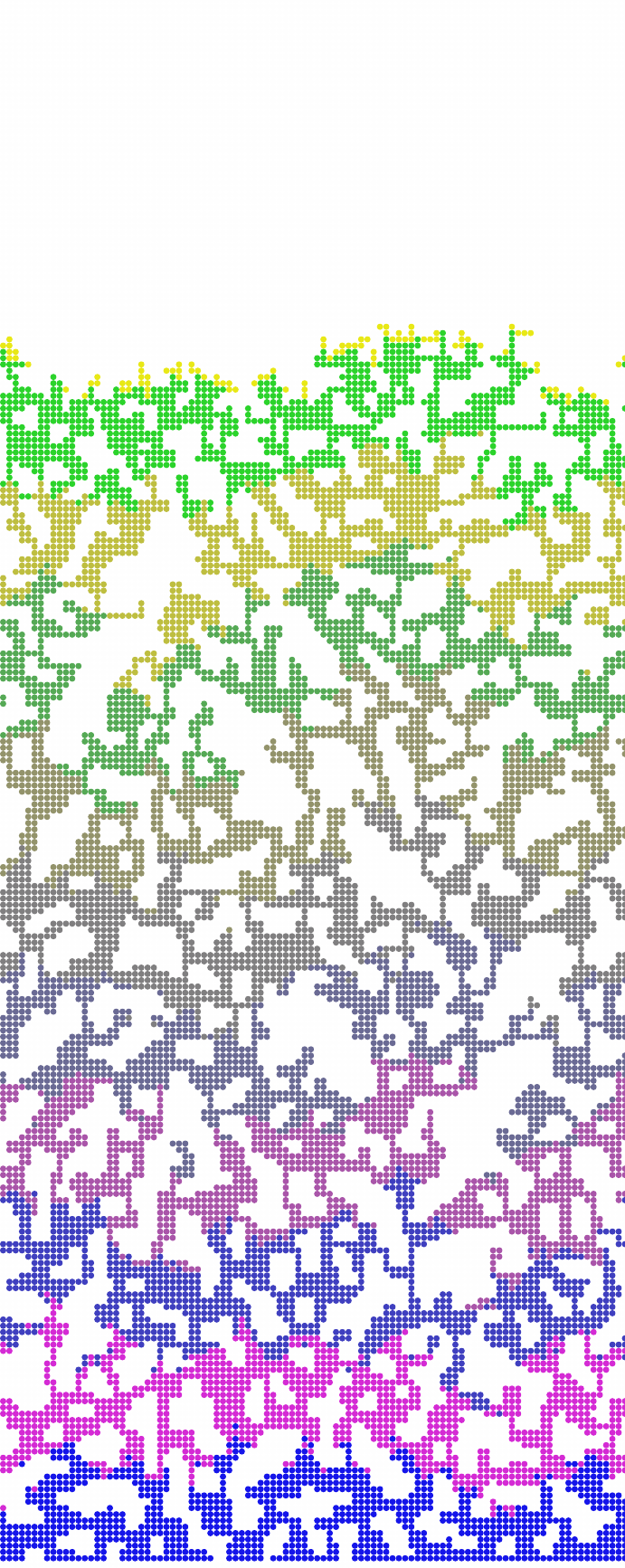} & \includegraphics[width=3cm]{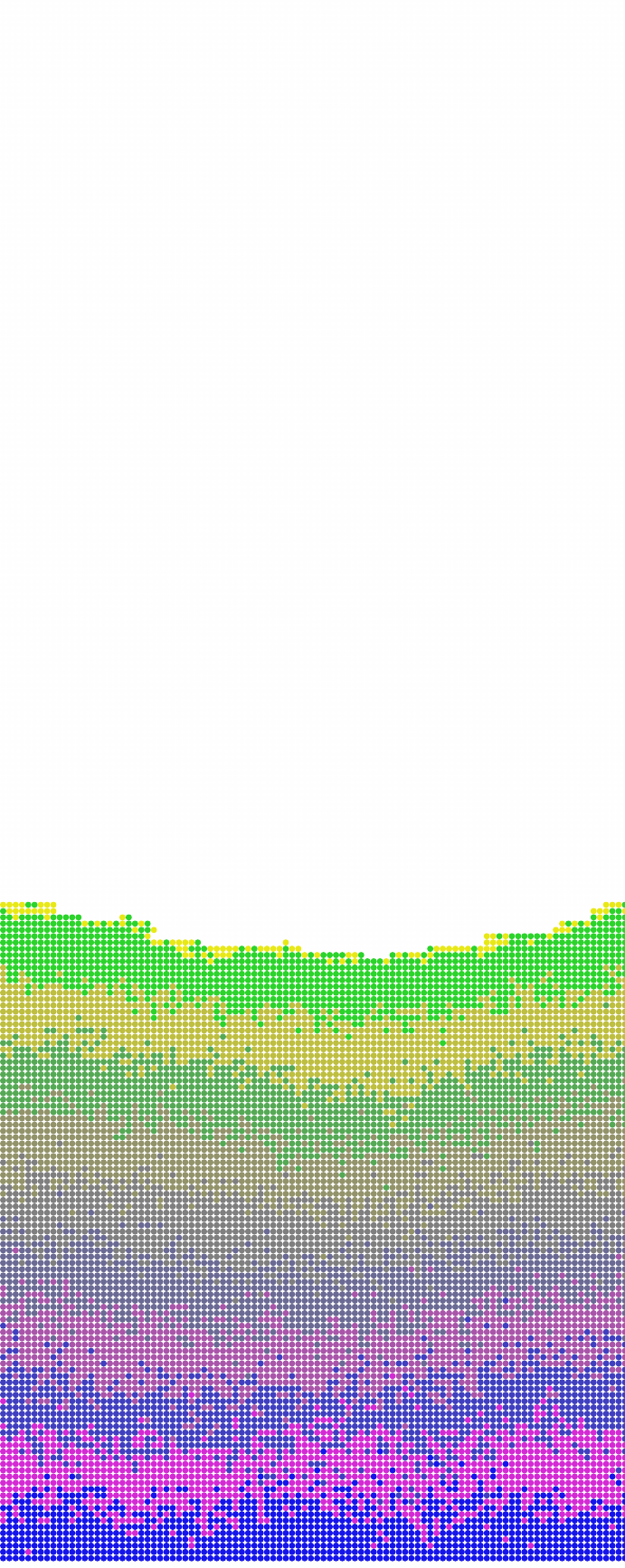} & \includegraphics[width=3cm]{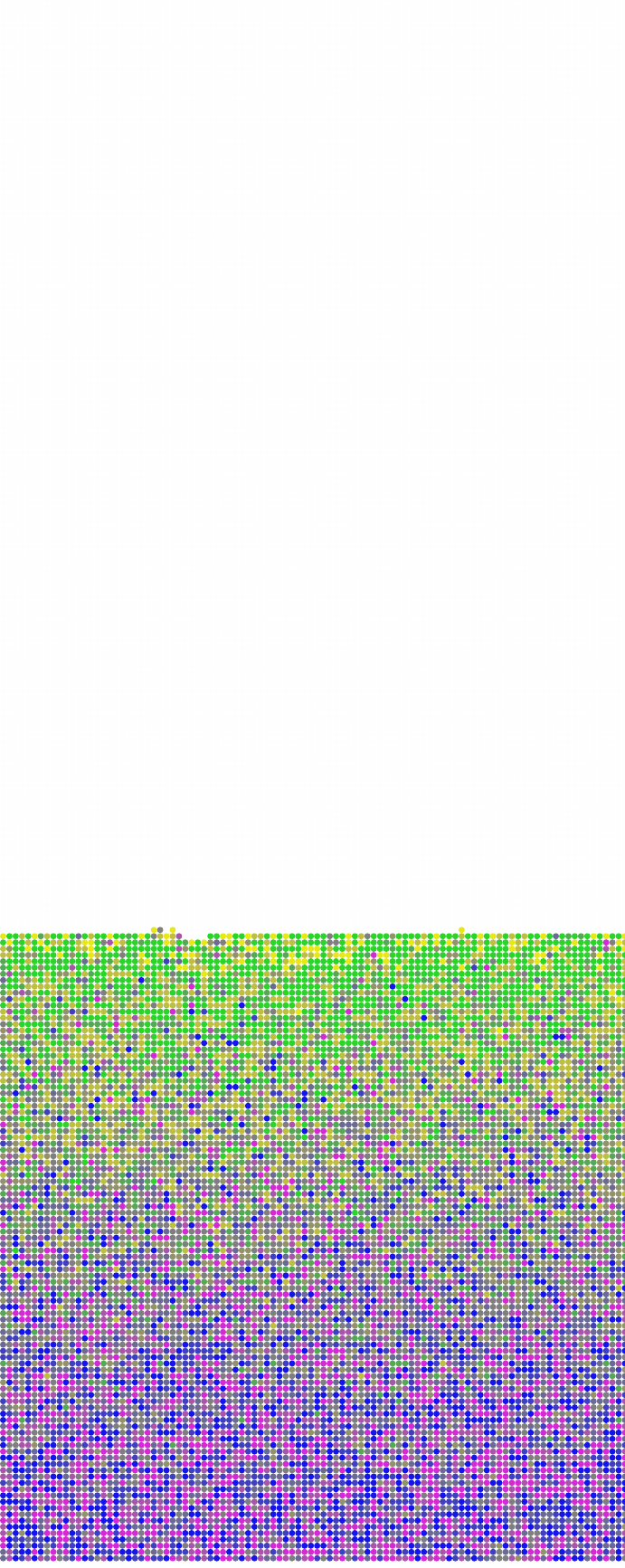}  \\
 a & b & c & d\\
 \end{tabular}
\caption{Snapshots of the different growth regimes a) BD ($\phi=3;F=1\times10^{15}$),  b) BDLR ($\phi=3;F=4\times10^{6}$), c) kinetic rough ($\phi=3;F=1\times10^{2}$), and d) layer-by-layer ($\phi=3;F=1$), for a system of 100 lattice sites and 100 deposited layers. Each color corresponds to 10 consecutively deposited layers.}
 \label{fig.snapshots}
 \end{center}
 \end{figure}

In figure~\ref{fig:roughness_exponent}(a) the evolution of the roughness is plotted for different values of the flux, with $\phi=3$ and $R_E=5$ kept fixed. After few deposited layers, different behaviors are observed and limiting regimes can be identified like, e.g., ballistic deposition at high fluxes and layer-by-layer at low deposition fluxes. For a characterization of the different growth regimes at various conditions the most commonly used measurement is the growth exponent $\beta$, given by,

    \begin{equation}
       W(t)\sim t^\beta.
       \label{eq:beta_exponent}
     \end{equation}   
In figure~\ref{fig:roughness_exponent}(b), we plot the value of $\beta$ as a function of $1/F$, for $R_E=5$ for different values of $\phi$. At fixed temperature (constant $\phi$), the inverse of the flux is a measurement of the mobility in the system. At low mobility, the ballistic deposition (BD) regime is recovered, characterized by shadowing of lower layers due to the overhanging of particles. As the mobility increases, particles relaxation is promoted and the roughness increases; we denote this regime as ballistic deposition with local relaxation (BDLR). As the density of vacancies vanish, a plateau in the roughness exponent as a function of the flux is observed, since small islands are formed, at the level of the first layer, and interlayer diffusion favors relaxation \cite{Krug2002} (at lower values of $\phi$ the plateau is suppressed). At lower fluxes, the size of islands increases, which makes interlayer diffusion less likely to occur. Consequently, island nucleation atop of existing islands leads to stratified growth and an increase in the roughness exponent. This growth mode, where mounds are formed, corresponds to the kinetic rough (KR) regime. For even lower values of flux (higher mobility), the system has enough time to relax between deposition events and a layer-by-layer growth (LBL) is observed \cite{Marmorkos1992}. Snapshots of various regimes are shown in figure~\ref{fig.snapshots}.

In the following subsections, these different transitions and their growth regimes are discussed. First, the transition to the thermal rough regime is addressed, then the one from BD to BDLR, and finally the transition from BDLR to kinetic rough growth. This section ends with the characterization of the transition from kinetic rough to layer-by-layer growth.

\subsection{Thermal roughening transition}\label{sec:TR}

\begin{figure}[t]
\begin{center}
\begin{tabular}{cc}
    \includegraphics[width=6.0cm]{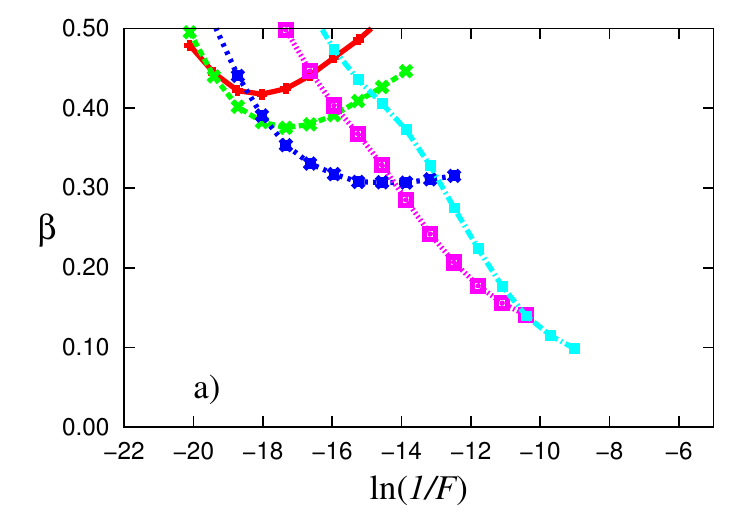} & \includegraphics[width=6.0cm]{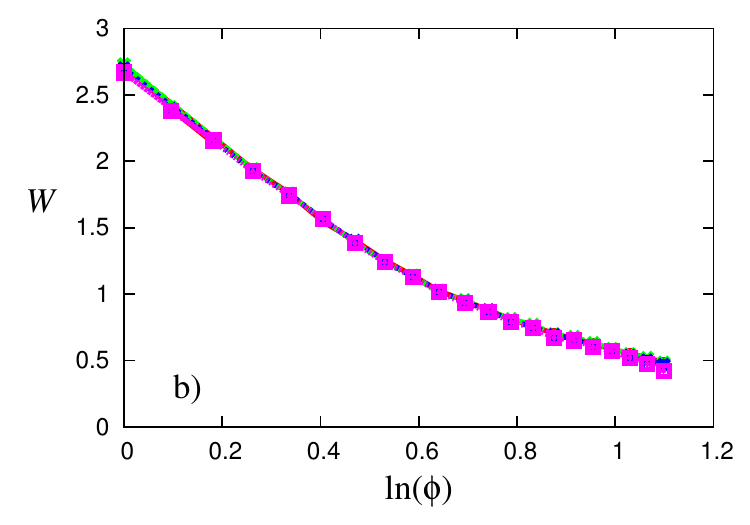}\\
\end{tabular}
    \caption{a) Growth exponent evolution as a function of $1/F$ for various $\phi$ values near $\phi_c$. From bottom to top we have $\phi=\{2, 1.8, 1.4, 1.2, 1.1\}$. b) Roughness as a function of the temperature for lattice sizes, from bottom to top, of 100, 200, 400, 1000 and 2000 on a previously deposited film of 10 ML.}
    \label{fig:exponent_phi_critic}
\end{center}
  \end{figure}

The roughening transition, or thermal roughening transition, as it is also known in the literature \cite{Lapujoulade1994}, intrinsically occurs under equilibrium conditions, though the remaining transitions referred to below occur away from equilibrium. To study this transition two different approaches are considered. We first consider its influence on the growth exponent $\beta$ (figure~\ref{fig:exponent_phi_critic}(a)) and we systematically study the behavior of an initially smooth surface at different temperatures (figure~\ref{fig:exponent_phi_critic}(b)).

The transition to the layer-by-layer regime is observed for a wide range of $\phi$ values, and occurs when the growth exponent becomes null. However, in figure~\ref{fig:exponent_phi_critic}(a), below a certain value of $\phi$, the growth exponent does not converge to zero, i.e., the transition to a layer-by-layer growth does not occur. In fact, the roughness exponent $\beta$ diverges at values of $\phi<2.0$, which can be interpreted as a critical $\phi_c$. To systematically study this equilibrium transition we consider an initially flat surface with 10 layers and let it evolve, for different values of $\phi$, to a constant roughness, which required $6\times10^4$ to $10^8$ time units. In figure~\ref{fig:exponent_phi_critic}(b), the roughness as a function of $\phi$ for different system sizes is shown. The plot shows a logarithmic increase of the roughness below the transition point $\phi_c$, as expected \cite{Lapujoulade1994}.

\subsection{BD to BDLR}

At high fluxes or low temperatures, two growth regimes are present, BD and BDLR (see figure~\ref{fig.snapshots}~(a)~and~(b), respectively), which differ by the presence of a partial restructuring during growth in the BDLR regime. However, this mobility is very low and particles mainly hop to nearby sites of higher coordination, i.e., sites with processes of higher activation barriers according to equation~(\ref{eq:arrhenius}).
 
\begin{figure}[t]
\begin{center}
\begin{tabular}{cc}
    \includegraphics[width=6.0cm]{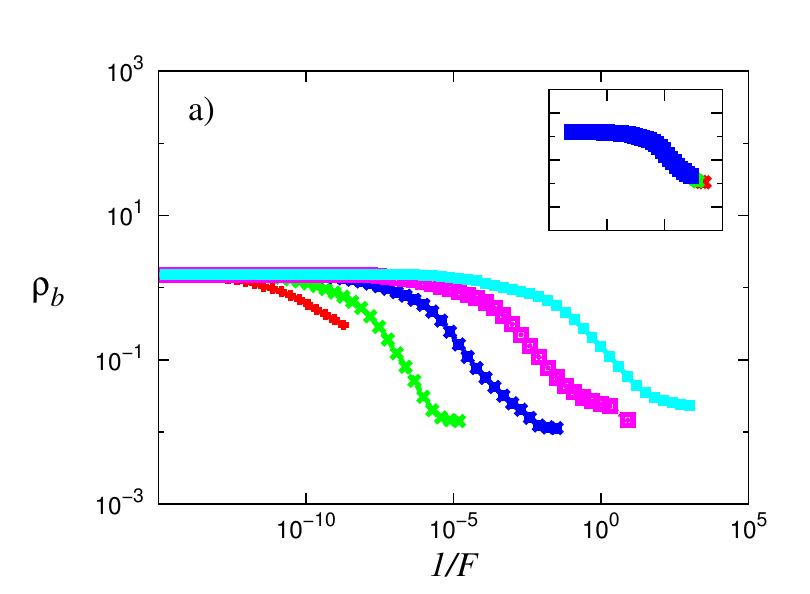} & \includegraphics[width=6.0cm]{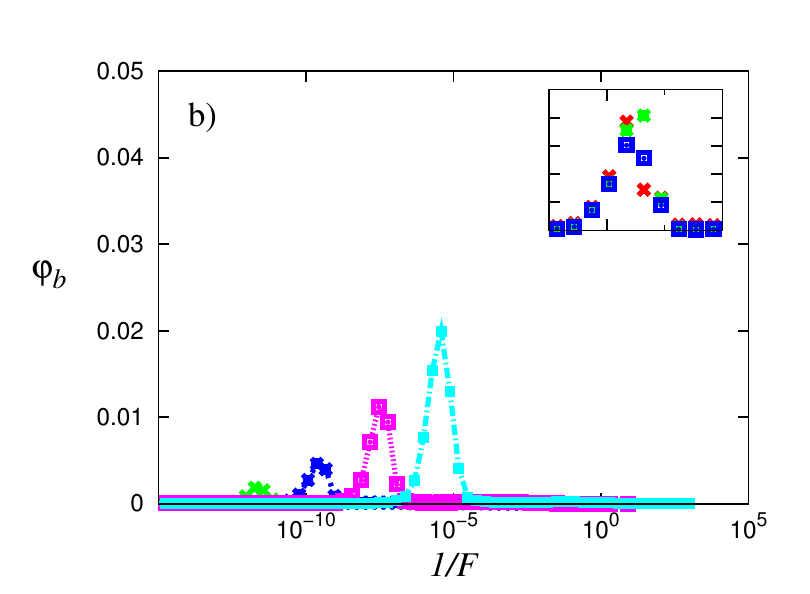}\\
\end{tabular}
    \caption{Density of dangling bonds (a), $\rho_{b}$, and its variance, $\varphi_{b}$, (b) as a function of $1/F$ for different values of $\phi$ after 100 deposited layers. From left to right, $\phi=\{1,2,3,4,5\}$. The insets are at $\phi=3$ and for system sizes of L=$\{400,1600,3200\}$.}
    \label{fig:roughness_bonddensity}
\end{center}
  \end{figure}

Despite similar behavior of the roughness for both regimes, the snapshots in figure~\ref{fig.snapshots} reveal different bulk structures. Therefore, the typical analysis based on the roughness exponent \cite{Barabasi1995} does not provide a way to differentiate between both regimes. Since the average coordination is higher in the BDLR regime, in figure~\ref{fig:roughness_bonddensity}(a) we analyze the behavior of the density of dangling bonds (bonds to perimeter sites) given by,

    \begin{equation}
       \rho_{b}=\frac{N_{b}}{LN_{l}},
       \label{eq:density_bond}
     \end{equation} 
where $N_{b}$ is the number of dangling bonds, $L$ is the linear size of the system, and $N_{l}$ is the number of deposited layers.

In figure~\ref{fig:roughness_bonddensity}(a) a constant density of dangling bonds is observed for higher fluxes which decreases at lower fluxes. The point where the density of dangling bonds starts to decrease, depends on the mobility of the particles. In the thermodynamic limit one expects the restructuring, at finite values of the flux, to relentlessly lead the bulk of the deposit to a more compact, layer-by-layer morphology. However, our interest focuses on the roughness  and the number of dangling bonds at the interface, which represent quantities that depend on the kinetic evolution of the growing interface. In a sense, the impact of mobility provides a measure of deviation from the extreme case of BD to BDLR. Consequently, the characterization of the transition is of a more subtle nature, based on the number of dangling bonds, while the growth exponent remains unaltered. For computational efficiency, we instead count the dangling bonds even in the bulk, since simulations are always carried over a finite number of (time) steps. At sufficiently low mobility, one does not expect substantial restructuring of the bulk, but accounting for these bonds provides much better statistics. An interesting behavior is found for the variance of the density of dangling bonds $\varphi_{b}$ in figure~\ref{fig:roughness_bonddensity}(b), given by
 
  \begin{equation}
   \varphi_{b}=\langle(\rho_{b}-\langle\rho_{b}\rangle)^2\rangle.
   \label{eq:second_density_bond}
  \end{equation} 
A maximum is observed, which can be taken to identify the transition from BD to BDLR. A non-vanishing maximum with the system size (inset of figure~\ref{fig:roughness_bonddensity}(b)) is a sign of a discontinuous transition in the thermodynamic limit \cite{Ferrenberg1998a}. Besides, the peak for the second moment shows a sharper discontinuity at the transition as particle mobility decreases.

\subsection{BDLR to kinetic rough}\label{subsec:BDLR_rough}

\begin{figure}[t]
\begin{center}
\begin{tabular}{cc}
    \includegraphics[width=6.0cm]{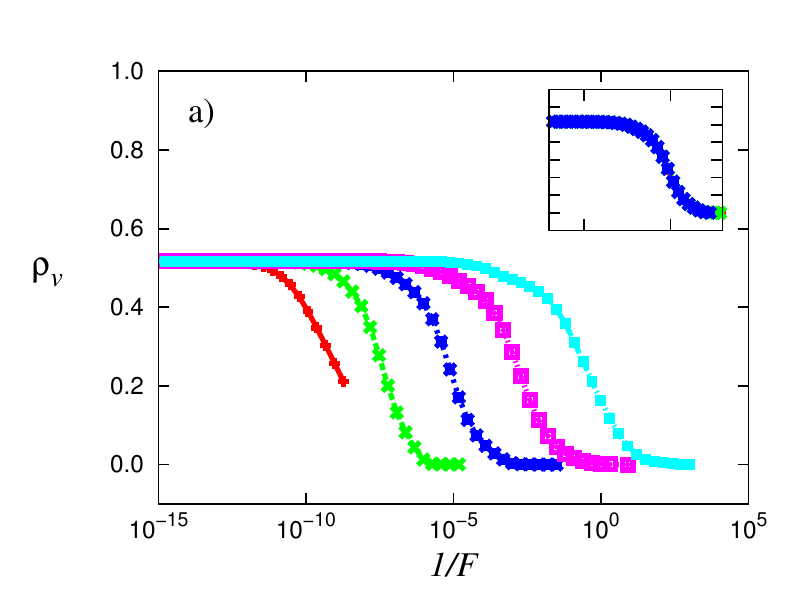} & \includegraphics[width=6.0cm]{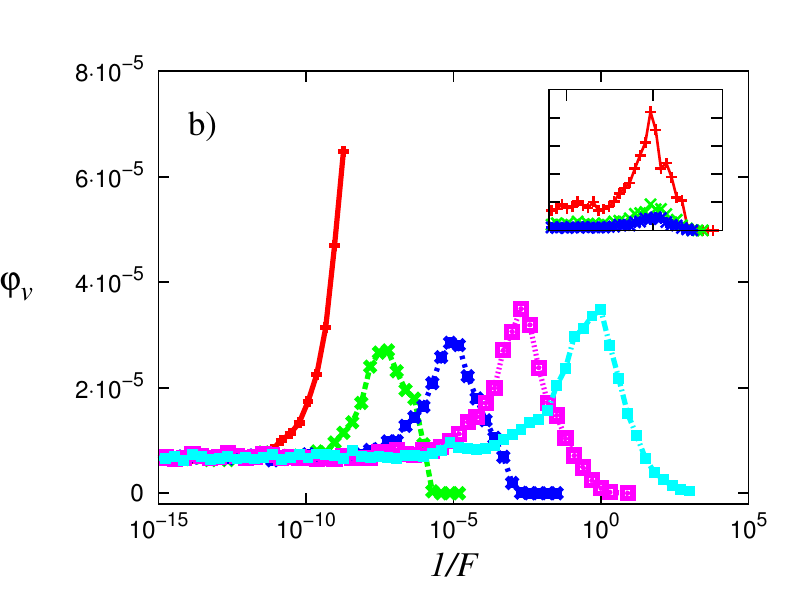}\\
\end{tabular}
    \caption{The density of vacancies (a), $\rho_{v}$, and its variance (b) as a function of $1/F$ for different values of $\phi$ after 100 deposited layers. From left to right, $\phi=\{1,2,3,4,5\}$. The insets are at $\phi=3$ and for different system sizes of L=$\{400,1600,3200\}$.}
    \label{fig:roughness_vacdensity}
\end{center}
  \end{figure}

As the flux lowers, particle diffusion becomes more relevant at a given temperature and particles placed on overhanging positions hop to more stable (higher coordination) positions, in part due to a larger relaxation time between consecutive deposition events. This behavior reduces the shadowing effect characteristic of the BD and BDLR regimes and, consequently, an increase in the roughness is observed (see figure~\ref{fig:roughness_exponent}(b)). Since the number of overhanging particles is reduced, the density of vacancies in the film also decreases. We measure this density, given by,

  \begin{equation}
  \rho_{v}=1-\frac{N_{l}}{\langle h\rangle}, 
   \label{eq:density_vacancies}
  \end{equation}
where $N_{l}$ is the number of deposited layers plotted as a function of $1/F$ in figure~\ref{fig:roughness_vacdensity}(a). A transition on $\rho_{v}$ is found, where it has a value of $1/2$, characteristic of the BD regime, and vanishes in the kinetic rough regime. As previously stated, the density of vacancies in BDLR does not significantly differ from that of BD, due to shadowing. Figure~\ref{fig:roughness_vacdensity}(b) shows the variance of the density of vacancies
  \begin{equation}
   \varphi_{v}=\langle(\rho_{v}-\langle\rho_{v}\rangle)^2\rangle,
   \label{eq:second_density_vacancies}
  \end{equation}
where $\langle...\rangle$ represents an average over samples. A maximum is observed, at the transition from BDLR to kinetic rough growth. However, the maximum is expected to vanish in the thermodynamic limit, a sign of a continuous transition (inset of figure~\ref{fig:roughness_vacdensity}(b)) \cite{Ferrenberg1998a}. A snapshot of the kinetic rough regime is shown in figure~\ref{fig.snapshots}(c).

\subsection{Kinetic rough to layer-by-layer}\label{sec:LBL}

Another relevant transition reproduced by this model is the kinetic roughening one between layer-by-layer and kinetic rough growth. The transition point can be computed by extrapolating in figure~\ref{fig:roughness_exponent}(b) the flux $F$, when the growth exponent, $\beta$, goes to zero. A snapshot of the layer-by-layer regime is in figure~\ref{fig.snapshots}(c)~and~\ref{fig.snapshots}(d).

  \begin{figure}[t]
\begin{center}
\begin{tabular}{cc}
    \includegraphics[width=6.0cm]{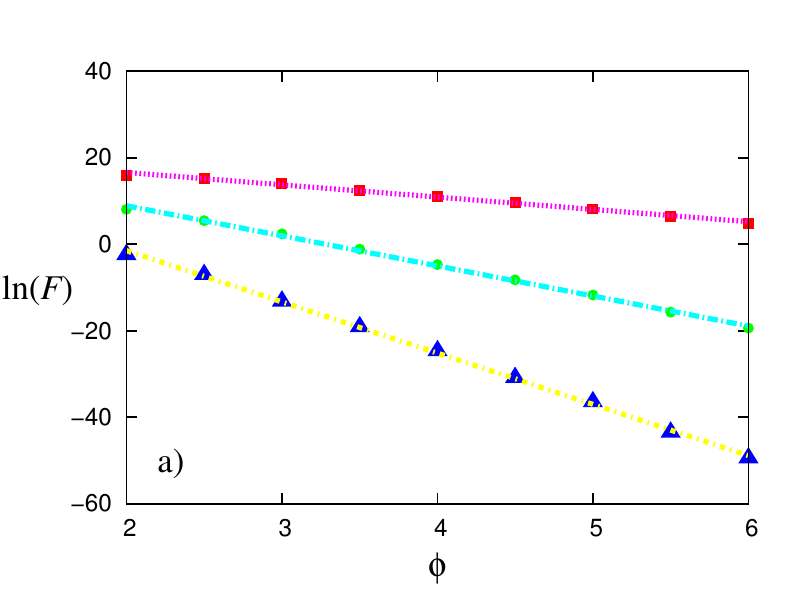} & \includegraphics[width=6.0cm]{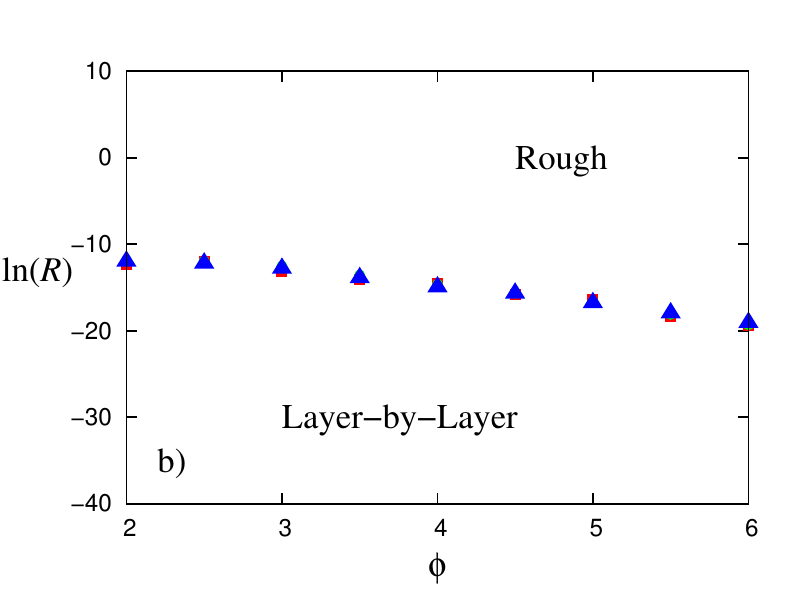}\\
\end{tabular}
    \caption{Growth mode diagram with the kinetic roughening transition between layer-by-layer and kinetic rough growth. a) The flux as a function of $\phi$ for values, from bottom to top, of $R_E=10$, $R_E=5$, and $R_E=1$. b) The ratio $R$ as a function of $\phi$ where triangles represent $R_E=10$, circles represent $R_E=5$, and squares represent $R_E=1$.}
    \label{fig:phase_diagram}
\end{center}
  \end{figure}

In figure~\ref{fig:phase_diagram}(a), the transition points from kinetic rough to layer-by-layer growth can be observed. In agreement with previous work \cite{Marmorkos1992, Marmorkos1990}, the transition has an Arrhenius dependency of the flux. For different values of $R_E$ (figure~\ref{fig:phase_diagram}(a)), different slopes of the Arrhenius exponential function, representing the kinetic roughening transition, are observed. A ratio $R$ is defined to relate the flux of particles $F$ and the terrace diffusion rate $\nu_{d}$,
 \begin{equation}
   R=\frac{F}{\nu_{d}}.
   \label{eq.ratio}
 \end{equation} 
The plot of $\ln(R)$ as a function of $\phi$, figure~\ref{fig:phase_diagram}(b), shows no influence from the various $R_E$ values, which represents, within the limits of the present study, a material independent property. Unlike the previous cases, for this transition a finite-size study has not been considered since the layer-by-layer regime represents a transient regime, which, at long timescales or large system sizes, crosses over to the kinetic rough regime \cite{Chatraphorn2002}.

A relation can now be established between the thermal roughening temperature $T_{tr}$ and the one from the kinetic roughening $T_{kr}$. This can be inferred from figure~\ref{fig:phase_diagram}(b), and the thermal roughening transition relation $\phi_{tr}=J_{pp}/T_{tr}$ as

 \begin{equation}
   F_c=A\nu_{d}\exp{\left(-\frac{\alpha \phi_{tr} T_{tr}}{T_{kr}}\right)},
   \label{eq.scaling_function}
 \end{equation} 
where $A$ and $\alpha$ are independent of the temperature and flux, so that they solely depend on $R_E$. This equation fits the plots of figure~\ref{fig:phase_diagram}(a), where the Arrhenius dependency of the flux over the temperature is observed. To reduce the dependency on $R_E$ in equation~(\ref{eq.scaling_function}), the ratio $R$ (equation~(\ref{eq.ratio})) is used, leading to the relation,

 \begin{equation}
   R=B\exp{\left(-c\frac{T_{tr}}{T_{kr}}\right)},
   \label{eq.scaling_function_fin}
 \end{equation} 
where $B$ and $c$ are constants. In figure~\ref{fig:phase_diagram}(b) are the same lines of figure~\ref{fig:phase_diagram}(a) with the proposed relation.

\subsection{Growth mode diagram}

\begin{figure}[t]
\begin{center}
    \includegraphics[width=6.0cm]{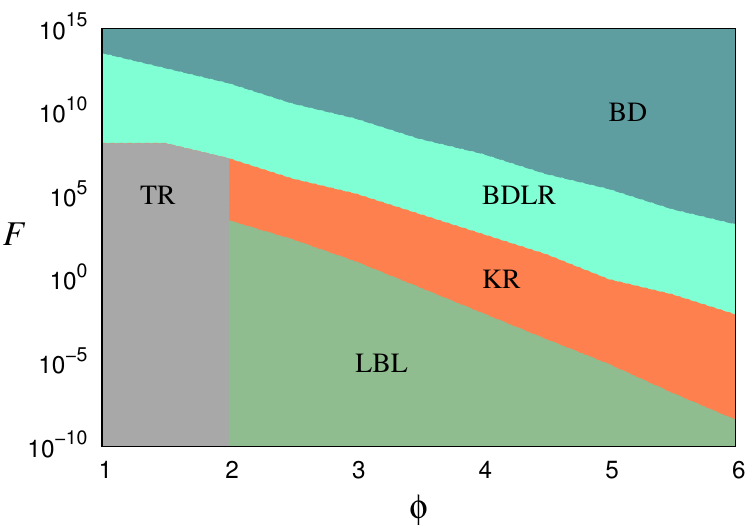}\\
\end{center}
    \caption{Final growth mode diagram, where TR and KR are the thermal rough regime and kinetic rough regime respectively.}
    \label{fig:phase_diagram_final}
  \end{figure}

The results are summarized in figure~\ref{fig:phase_diagram_final}. The diagram represents, for different pairs of $\phi$ and $F$, the corresponding growth regime. BD/BDLR and BDLR/KR transition lines were picked from figures~\ref{fig:roughness_bonddensity}(b)~and~\ref{fig:roughness_vacdensity}(b), respectively. For the KR/LBL and thermal roughening transition results from sections~\ref{sec:LBL}~and~\ref{sec:TR} were considered. The diagram of figure~\ref{fig:phase_diagram_final} represents the observed growth modes and the transition lines between them.

\section{Concluding remarks}\label{conclusions}
We presented a KMC model to study the influence of both the flux and temperature on the epitaxial growth. We simulate an (1+1)-dimensional lattice mode with nearest-neighbor interactions that includes deposition and thermally activated processes. Despite the simplicity of the model, known regimes are reproduced like, e.g., ballistic deposition, kinetic rough, layer-by-layer, and thermal rough growth. Additionally, a new regime between BD and KR is identified, the ballistic deposition with local relaxation.

We studied the transitions between the various regimes with emphasis on the BD/BDLR and BDLR/KR ones, where the traditional surface analysis has to be complemented by other properties like the number of dangling bonds or the number of vacancies. The transition KR/LBL was also studied and the Arrhenius relation between temperature and flux confirmed. Furthermore, we showed a relation between the thermal and kinetic roughening transitions.
 
A more detailed study of the scaling properties is left open as well as the straightforward extension to a study in 2+1 dimensions. 

 \section*{References}
  \bibliography{clean_Films_Paper}

\end{document}